\documentstyle[11pt,epsfig]{article}
\begin{document}

 \newcommand \be {\begin{equation}}
\newcommand \ee {\end{equation}}
 \newcommand \ba {\begin{eqnarray}}
\newcommand \ea {\end{eqnarray}}
 \newcommand \bea {\begin{eqnarray}}
\newcommand \eea {\end{eqnarray}}

\newcommand{\siml}{\stackrel{<}{\sim}}
\def \p {\vec \phi({\bbox x})}
\def \dk { {\int  \frac{d^D{{{\bbox k}}}{(2\pi)^D}} }}
\def \(({\left(}
\def \)){\right)}
\def \[[{\left[}
\def \]]{\right]}
\def \cZ{{\cal Z}}
\def \eps{\epsilon}

\title{\bf UNIVERSALITY CLASSES FOR EXTREME VALUE STATISTICS}

\vskip 3 true cm

\author{Jean-Philippe Bouchaud$^1$ and Marc M\'ezard$^2$}
\vskip 1 true cm

\date{\it $^1$ Service de Physique de l'\'Etat Condens\'e,
 Centre d'\'etudes de Saclay, \\ Orme des Merisiers, 
91191 Gif-sur-Yvette Cedex, France \\ 
$^2$ Laboratoire de Physique Th\'eorique de l'Ecole Normale Sup\'erieure
 \footnote {Unit\'e propre du CNRS,  associ\'ee
 \`a\ l'Ecole
 Normale Sup\'erieure et \`a\ l'Universit\'e de Paris Sud} , \\
24 rue
 Lhomond, 75231 Paris Cedex 05, France }


\maketitle

\begin{abstract}
{The low temperature physics of disordered systems is governed by the
statistics of extremely low energy states. It is thus rather important to 
discuss the possible universality classes for extreme value statistics. We compare the usual probabilistic classification to the results of the replica approach. We show in detail that one class of independent variables corresponds exactly to the so-called {\it one step replica symmetry breaking} solution in the replica language. This universality class holds if the correlations are sufficiently weak. We discuss the relation between the statistics of extremes and the problem of Burgers turbulence in decay.}
\end{abstract}

\vskip 0.5cm

LPTENS preprint 97/XX

\vskip 0.5cm

\noindent Electronic addresses : 
bouchaud@amoco.saclay.cea.fr, mezard@physique.ens.fr


\section{Introduction}

The replica method is one of the very few general analytical methods to investigate 
disordered systems \cite{MPV}. Although the physical
meaning of Parisi's `replica symmetry breaking' ({\sc rsb}) scheme
needed to obtain the correct low temperature solution of various random models
 has already been discussed on several occasions \cite{MPV}, its precise
relation with the so-called extreme value statistics \cite{Gumbel,Galambos} (and therefore its scope and
limitations) was not previously clearly established. 
That such a relation should
exist is however intuitively obvious: at low temperatures, a disordered system will
preferentially  occupy its low energy states, which are random variables because of
the disordered  nature of the problem. The statistics of the free-energy (or of
other observables, such as energy {\it barriers} \cite{Rammal}) will thus reflect the statistics of these low energy (extreme)
states. It is well known in probability theory that extreme value
statistics can be classified into different universality classes \cite{Gumbel,Galambos}.
Conversely, the {\sc rsb} scheme has shown the existence of at
least two broad classes of systems, those with a first order, `one step' {\sc rsb}
and those with continuous {\sc rsb}.

It is easy to identify the `one step {\sc rsb}' class with one particular 
universality class of extreme value statistics, i.e. the Gumbel class,
which concerns the minimum of continuous variables which are unbounded
but have a distribution decaying faster than any power at $-\infty$.
The simplest representative of this class is the Random 
Energy Model ({\sc rem}) \cite{REM}, where the energy
states are independent (but not necessarily Gaussian distributed).
An interesting point is that the {\sc rem} can be given a spatial
structure, for  which the replica theory still provides the {\it exact} solution. This
spatial {\sc rem} is in turn connected, in one dimension, to the problem of
decaying Burgers' turbulence \cite{Burgers} in which an infinitely compressible
fluid evolves from random initial conditions. Exact results
for the velocity correlations at large times in Burgers' turbulence
have been obtained long ago by Kida \cite{Kida}. We shall show that these results
coincide with those of the replica method, the underlying reason being that the late
stage of turbulence decay is governed by the extreme values
 for the integral of  the initial velocity field. 

It is less easy to identify the other universality classes of extreme
value statistics. There should be at least two types of generalizations.
One type still concerns {\it independent} random variables but with either
power law decay of the distribution (in which case there is 
a priori no replica formalism), or bounded random variables (the Weibull
distribution of extremes), which does not
seem to correspond to any known {\sc rsb} scheme. The other type concerns 
{\it correlated} variables, for which the only results known to us
are actually derived in the framework of replicas: those are cases of full {\sc rsb}, which describe random variables
with a certain (hierarchical) type of correlations.

These universality classes are the counterpart for extremes of random
variables to the usual universality classes studied in the framework
of sums of random variables. Taking the well known example of random walks or
polymers, the usual random walk, or ideal polymer, is  described
asymptotically by
the Gaussian central limit theorem, while the addition of independent variables with infinite variance leads to new universality classes (L\'evy sums) \cite{BG}. The introduction of long-range correlations like self avoidance also leads to totally new universality classes \cite{BG}. We wish here to take a
first step in an analogous 
categorization for extreme values, which appear naturally in
disordered systems at low temperatures.

\section{Extreme value statistics}

\subsection{Scaling regime}

We  start by recalling standard results of extreme values statistics, in order
to set the 
stage for the following discussions. Consider $M$ independent, identically 
distributed random variables $E_i$, $i=1,...,M$ (`energies'), such that the
probability distribution decays for $E_i \to -\infty$ faster than any power-law: \be
P(E) \sim \frac{A}{|E|^\alpha} \exp[-B |E|^\delta] \qquad B,\delta > 0
\ ; \ \ \ E \to -\infty
\label{probainit}
\ee
We are interested in the statistics of the lowest energy state 
$E^*=\min\{E_1,...E_M\}$ 
for large $M$. Defining ${\cal P}_<(E)$ as the repartition function of $E$:
\be
 {\cal P}_<(E)=\int_{-\infty}^E dE' P(E')
\ee
one can express the distribution $P_M$ of $E^*$ as:
\be
P_M(E^*)= M P(E^*) [1-{\cal P}_<(E^*)]^{M-1} = - \frac{d}{dE^*} [{\cal P}_>(E^*)]^{M}
\ee
For large $M$, the minimum $E^*$ will be negative and large, so that:
\be
[1 -{\cal P}_<(E^*)]^M \simeq \exp[-M {\cal P}_<(E^*)]
\ee
The repartition function of $E^*$ thus becomes very small when $E^*$
is smaller than
the characteristic value of the energy $E_c$ defined by
$M {\cal P}_<(E_c)=1$. To logarithmic accuracy, this gives in the
case of the distribution (\ref{probainit}):
\be
E_c \simeq - \left(\frac{\log M}{B}\right)^{1/\delta} \label{ec}
\ee
Defining now $E^*=E_c+\epsilon$, with $\epsilon \ll |E_c|$, one has, to first order:
\be
[1-{\cal P}_<(E^*)]^{M} \simeq \exp[-\exp(B\delta |E_c|^{\delta-1}\epsilon)]
\label{ecprime} \ee
Finally, introducing the rescaled variable $u=B\delta |E_c|^{\delta-1}\ \epsilon$, one finds
that the rescaled variable obeys, for large $M$, a universal `Gumbel' distribution \cite{Gumbel,Galambos}
$P^*$, independent of the coefficients $A,B$ and exponents $\alpha,\delta$:
\be
P^*(u) = \exp(u-\exp u) \label{gumbel}
\ee
A very important property, which we shall emphasize later on, is that $P^*(u)$ 
vanishes exponentially for $u \to -\infty$ (and much faster still for $u \to
+\infty$). The maximum of $P^*(u)$ occurs at $u=0$, meaning that $E_c$ is actually
the most probable value for the extreme energy. Finally, as in any `central' limit
theorem, this behaviour is only valid in the region where the deviation $\epsilon$
from $E_c$ is of the order of $E_c^{1-\delta}/B$, which goes to zero with $M$ if
$\delta > 1$ and diverges otherwise. The {\it relative} fluctuations $\epsilon/E_c$,
however, are always of order $1/\log M$.

\subsection{The large M limit and the Random Energy Model}

Let us now consider the following partition function:
\be
{\cal Z} = \sum_{i=1}^M z_i \qquad z_i=\exp[-\frac{E_i}{T}]
\ee
where the $E_i$ are distributed as in (\ref{probainit}). This is a
slight generalization of Derrida's original {\sc rem}, initially introduced
with  a purely Gaussian distribution ($\delta=2$). Obviously,
the independent variables $z_i$, 
are large when $E_i$ is large and negative. In the scaling region defined above, due to the exponential tail
of (\ref{gumbel}), the distribution of $z$ decays for large $z$ as a power-law: 
\be
P(z) \propto z^{-1-\mu} 
\qquad ; \qquad z \to \infty
 \label{power}
\ee
where $E_c$ is the most probable ground state energy of the system, given by 
Eq. (\ref{ec}), and 
\be
\mu =T B \delta |E_c|^{\delta-1}  \ .
\label{mumu}
\ee
The partition sum 
$\cal Z$ behaves very differently in the region $\mu<1$, where the 
average
value of $z$ diverges and thus only a small number of terms (those of order $M^{1/\mu}$) 
contribute to $\cal Z$, and in the region $\mu > 1$, where all the $M$ terms give a
(small) contribution to $\cal Z$. This means that for 
\be
T_c = \frac{1}{B\delta E_c^{\delta-1}}\label{onze}
\ee
for which $\mu=1$, the probability measure concentrates onto
 a finite number of states, corresponding to the glass transition in
these models. In the random energy model, $M$ is the number of states $M=2^N$. 
In order to have an extensive ground state energy 
($E_c \propto N$) and $T_c$ finite in the large $N$ limit,
 one should choose (see Eqs. (\ref{ec},\ref{onze})) $B = N^{1-\delta}$. For $\delta=2$, this indeed coincides with the 
usual scaling of the energies in the {\sc rem}. 

Let us now study the statistics of the weights $p_i \equiv z_i/{\cal Z}$ in the 
glassy
region $T < T_c$. Since:
\be
w_i = \frac{z_i}{z_i + {\cal Z}'}
\ee
where ${\cal Z}'=\sum_{k (\ne i)} z_k$ is independent of $z_i$ (and of order 
$M^{1/\mu}$), one readily finds that \footnote{We denote as $P(.)$ the probability 
density of the variable appearing in the parenthesis; hopefully there is not
ambiguity in the following.}:
\be
P(w) = \frac{{\cal Z}'}{(1-w)^2} P\((z=\frac{{\cal Z}'w}{1-w}\))  
\ee
For $w_i$ to be non zero in the large $M$ limit, $z_i$ has to be large. In
that region   one can 
 use the
asymptotic form (\ref{power}) for $P(z)$, giving:
\be
P(w) = \frac{C}{M} (1-w)^{\mu-1} w^{-1-\mu} \qquad w \gg M^{-1/\mu} \label{Pw}
\ee
where $C$ is a constant fixed by the condition $M\int_0^1 dw w P(w)\equiv 1$. From
 this probability distribution of each weight, one
can deduce the  moments  $Y_k \equiv 
\overline{ \sum_i w_i^k} $, which characterize 
to what extent the measure concentrates onto a few states: if all weights 
are of the same order of magnitude, then $Y_k \sim M^{1-k} \to 0$ for $k > 1$; while if 
only a finite number of weights contribute, the moments $Y_k$ remain finite 
when $M \to \infty$. In the present case, one finds, for $\mu < 1$,
\be
Y_k = M \int_0^1 dw w^k P(w)=\frac{\Gamma[k-\mu]}{\Gamma[k]\Gamma[1-\mu]} \qquad (k > \mu)
\label{Ykdirect}
\ee
(see also \cite{Derrida}). Since $\mu = T/T_c$, one finds that $Y_2$ goes 
linearly to zero for $T \to T_c$, and that $Y_k = 1 - (\Gamma'[k]-\Gamma'[1])/\Gamma[k])T/T_c$ for $T \to 0$. 

Finally, the average  energy per degree of freedom of the system 
is constant throughout the low temperature phase ($T<T_c$) and given by 
\be
\overline E/N = E_c/N + <u> B\delta E_c^{\delta-1}/N 
\sim  - (\log 2)^{1/\delta} + O(1/N)\ ,
\label{enedirect}
\ee
 where the average $<u>$ is taken over the Gumbel distribution,
giving: $<u>=\Gamma'[1]$.

\section{The replica approach}
\subsection{The REM}
We shall now show how all these results can be recovered using the
replica method. We suppose that $\delta > 1$ (the case $\delta<1$ will
be discussed below) 
and introduce the characteristic function $g(\lambda)$ through:
\be
\int_{-\infty}^{\infty} dE P(E) \exp[-\lambda E] \equiv \exp[g(\lambda)]
\ee
Since $B=N^{1-\delta}$, this integral can be
computed at large $N$ with  a saddle-point method, which gives: 
\be
g(\lambda)= (\delta-1) N \left(\frac{\lambda}{\delta}\right)^{\frac{\delta}
{\delta-1}}
\ee

In the replica method we need to compute the moments of
the $\cZ$ distribution:
\be
\overline{{\cal Z}^n} = \overline{ \sum_{i_1,i_2,...i_n} 
z_{i_1} z_{i_2} ... z_{i_n}}
\equiv \overline{ \sum_{i_1,i_2,...i_n} \exp\left[-\frac{1}{T} \sum_{i} 
E_i \sum_{a=1}^n \delta_{i,i_a} \right] }
\ee
The averaging over the $E_i$ gives:
\be
\overline{{\cal Z}^n} = \sum_{i_1,i_2,...i_n} \exp[\sum_i 
g(\frac{1}{T} \sum_{a=1}^n \delta_{i,i_a})]
\label{Zn}
\ee
The point now is to understand which configurations of $\{i_1,i_2,...i_n\}$ will 
dominate
the above sum when $N \to \infty$ (and $n \to 0$).
 The simplest Ansatz, corresponding to the largest phase space volume, assumes that 
all $i_a$ are different, 
leading to:
\be
\overline{{\cal Z}^n} = M(M-1)...(M-n+1) \exp[n g(\frac{1}{T})] \simeq 
\exp[n (\log M
+g(\frac{1}{T}))]
\ee
Taking $n \to 0$, one thus finds that the free energy per degree of freedom 
$f = -\frac{T}{N} \overline{ \log{\cal Z} }$ takes the value:
\be
f =f_0 \equiv -T \log 2 -  (\delta-1) \delta^{-\frac{\delta}
{\delta-1}} T^{-\frac{1}{\delta-1}} \label{f0}
\ee
The entropy $s_0 = -df_0/dT$ is therefore equal to:
\be
s_0=\log(2)-(\delta T)^{-\frac{\delta} {\delta-1}}
\ee
and becomes negative below a critical temperature 
\be
T_c={1 \over \delta} \log(2)^{1-\delta \over \delta}
\ee
So this solution, called `replica symmetric'  (since all
replicas $i_a$ play a symmetric role), has to be modified
in the low temperature phase. The correct configurations which
dominate the sum (\ref{Zn}) at $T<T_c$ are called `one step replica symmetry breaking'
 and are such that the $n$ replica indices $\{i_1,i_2,...i_n\}$
are grouped into $n/m$ groups of $m$ equal indices, which can be 
written after a proper relabelling:
\bea
i_1=i_2=...=i_m=k_1\\
i_{m+1}=i_{m+2}=...=i_{2m}=k_2\\
...\\
i_{n-m+1}=...=i_n=k_{n/m}
\eea
and now the indices $k_1,...,k_{n/m}$ are all different one from
the other. These configurations contribute
to $\overline{{\cal Z}^n}$ as:
\be
\overline{{\cal Z}^n} = M(M-1)...(M-n/m+1) \exp[{n \over m}
 g(\frac{m}{T})] {n! \over m!^{n/m}}
\ee
from which one immediately deduces:
\be
f(T)=f_0(T/m)
\ee
where $f_0$ is defined in Eq. (\ref{f0}). The extremum of this free energy with respect to $m$ is obtained
when
\be
{\partial f \over \partial m}=0=s_0(T/m)
\ee
which gives 
\be
 m={T \over T_c}=\mu
\ee
Note that this relation is independent of $\delta$.
Therefore this one step {\sc rsb} solution predicts that the
system freezes at the critical temperature $T_c$ which is
the temperature where the entropy $s_0$ vanishes. The energy density is
constant throughout the low temperature phase, and equals:
\be
e=f_0(T_c)=- (\log \ 2)^{1/\delta}
\ee
in agreement with the direct computation (\ref{enedirect}). Since the free-energy is
constant, the entropy of the whole low temperature phase is zero \cite{REM}.

It turns out that also the finer details, like the distribution of the weights of
the configurations which dominate the low temperature
measure, can be computed by this replica approach \cite{MPSTV}.
 By definition, the moments
$Y_k$ are equal to:
\bea
Y_k&=&\overline{ \sum_i \frac{z_i^k}{{\cal Z}^k} } =\lim_{n\to 0} \overline{ \sum_i z_i^k {\cal Z}^{n-k} } \\
&=& \lim_{n \to 0} {1 \over n(n-1)...(n-k+1)} \sum_{a_1,..,a_k}'
 \sum_{i_1,...,i_n} \overline{ z_{i_1}...z_{i_n} }
\prod_{j=1}^k \delta_{i_{a_1},i_{a_j}}
\eea
where the sum primed over the $a$'s runs from $1$ to $n$, with all $a$'s different. Owing to the 
structure of the {\sc rsb}, this means that one simply has to pick the  $k \leq m$ replica indices 
${a_1,...,a_k}$ in the same `group', for which there are $(m-1)...(m-k+1)$ possibilities once
$a_1$ has been chosen. Hence:  
\be
Y_k=\lim_{n \to 0} {n(m-1)...(m-k+1) \over n(n-1)...(n-k+1)}\overline{{\cal Z}^n}=  \frac{\Gamma[k-\mu]}{\Gamma[k]\Gamma[1-\mu]} 
\ee
in agreement with the direct computation (\ref{Ykdirect}).

\subsection{The {\sc rem} with $\delta < 1$: a first order transition}

\begin{figure}
\centerline{\hbox{\epsfig{figure=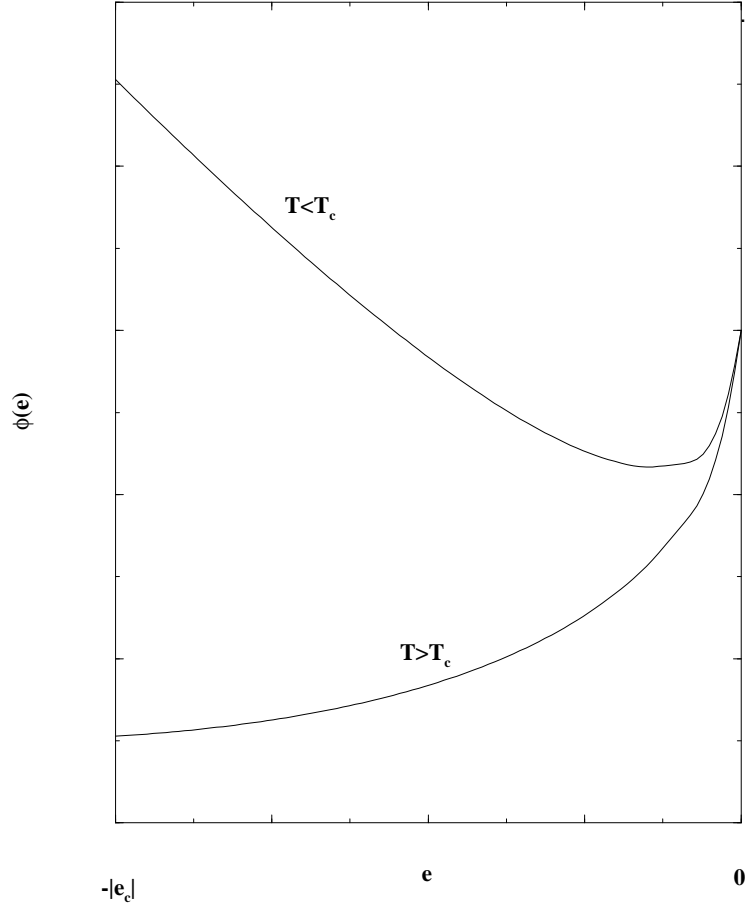,width=8cm}}}
\vskip 0.8cm
\caption{$\varphi(e)$, as defined in Eq.~(\protect\ref{part}), as a function of $e$ for different temperatures. The saddle point is thus at $e=0$ for $T>T_c$ 
and at $e=-|e_c|$ for $T<T_c$. There is no states (in the limit $N \to \infty$ beyond $-|e_c|$.}
\label{fig1}
\end{figure}

The above method fails when $\delta \leq 1$, which actually corresponds to a 
different universality
class from the point of view of critical phenomena, while the nature of the
low temperature phase leaves it in the same class as the systems 
with $\delta>1$, in agreement with the extreme value classification
which does not distinguish between $\delta>1$ or $\delta <1$.
 In order to study the transition, we use Derrida's original 
`microcanonical' method. Using the normalisation $B=N^{1-\delta}$, the partition function is equal to:
\be
{\cal Z} =  \int_{-e_c}^0 de \exp  N \varphi(e) \qquad \varphi(e)= \log 2 - |e|^\delta + \frac{|e|}{T}\label{part}
\ee 
where $e=E/N$,  and $e_c$ is the energy density beyond which there 
are no states
(for $N \to \infty$), i.e.: $2^N \exp(-N e_c^\delta) = 1$. As shown in Fig. (\ref{fig1}),
the integral is dominated 
either by $e=0$
or by $e=-e_c$, depending on the temperature. When 
$T>T_c=(\log 2)^{1-\delta/\delta}$, the free energy is equal to $-NT \log 2$, while
for $T<T_c$, the free energy  is equal to a constant $-N e_c=-N (\log 2)^{1/\delta}$.
The transition at $T_c$ is now a first order transition from the
thermodynamic point of view, with a jump in  the entropy.
This is in contrast with the usual case $\delta > 1$ where the transition
is thermodynamically of second order \footnote{Although there is a jump in the Edwards-Anderson order parameter \cite{REM}}.
\vskip 0.8cm

In the low temperature phase, only the neighbourhood of $E_c=-N|e_c|$ is of 
importance, and we get back to the universal Gumbel distribution since
the density of states is  still locally exponential:
\be
P(E=E_c+\epsilon) \propto \exp \frac{\mu \epsilon}{T} \qquad \mu \equiv 
 \frac{T}{T_c} \delta
\ee
The value of $\mu$ again determines the statistics of the weights, as above. 
Note that however that for $\delta < 1$, the value of the parameter $\mu$ (corresponding to the {\sc rsb} parameter) is smaller than $1$ at the transition $T=T_c^-$. Hence $Y_2$ is discontinuous at $T=T_c$, in contrast to the case $\delta > 1$.

\subsection{Physical interpretation of the replica solution}

The reason why replica symmetry must be broken in order to get sensible results
in this problem  is  rather 
clear. Since the distribution of the Boltzmann weights $z_i$ is a power law with an 
exponent $\mu < 1$ in the low temperature phase, all integer moments of $\cal Z$ (and
thus of $z$) are formally divergent, and are thus dominated by a cut-off for large
$z$ which has nothing to do with the value of the `typical' $z$'s, and hence of the
free-energy. Calculation based on a simple analytic continuation of the results
obtained for $n>1$ are thus bound to fail. The replica method with one step
{\sc rsb} manages to compute
$\langle z^m \rangle$ with $m=\mu$, which precisely picks up the contribution 
of the typical region
of $z$.(Smaller values of $m$ would be mostly sensitive to very small $z$, while larger $m$'s probe atypically large values of $z$.) The algebra corresponding to one step {\sc rsb} exactly
reproduces the extreme value statistics in the case of fastly decaying
distributions. In this respect, {\sc rsb} does not mean more than a `localisation'
of  weights onto a small subset of all configurations, in the sense that major
contribution to the partition function comes from a finite number of configurations
(i.e. all $Y_k >0$) \cite{MPSTV}. Actually, the quantities $Y_k$ were also introduced 
in the context of electron localisation in disordered potentials, and called
`participation ratios' \cite{Chamon}.

\section{A d-dimensional Random Energy Model}

In this section we want to study a generalized version of the {\sc rem},
where the energy levels are embedded in a euclidean space. Besides its
intrinsic interest as a model for a particle in a disordered environment, 
this problem turns out to be also directly relevant to the study of declining 
Burgers turbulence, 
as we shall discuss in detail in the next section. 

The model is defined as follows. To each point $x$ of a
(discretized) $d$-dimensional space, one assigns a potential energy
$E(\vec x)$ which is a random number picked up independently
on each point, from a distribution $P(E)$ the tail of which
is given by (\ref{probainit}). The total energy on this
point is the sum of a deterministic part, which we take for
instance equal to $\kappa x^2/2$, and this random contribution $E(\vec x)$.
This defines a certain energy landscape, to which we associate
a partition function ${\cal Z}$ as:
\be
{\cal Z} = \int d^d x \exp\((-{V(\vec x) \over T}\)) \ \ , 
\ \ V(\vec x)\equiv {\kappa x^2\over 2}+E(\vec x)\label{Z}
\ee
Here we adopt a continuum notation but an ultraviolet cutoff (lattice spacing)
is implicitely assumed when necessary. The role of the deterministic part
proportional to $\kappa$ is twofold. First of all it allows one to define 
a topology in the space of the points $\vec x$ (The limit $\kappa=0$coincides with the {\sc rem}, the fact that the points sit in
a $d$-dimensional space being irrelevant). For this purpose
the deterministic part could be rather arbitrary, and indeed one can solve
the problem with a more general deterministic energy. As we shall need a
quadratic term later on, and in order to keep the presentation simple, we 
restrict to this particular case. Second, the presence of this confining term 
allows one to deal with this model without the need of introducing a finite box. 

This model with $d=1$ was in fact introduced
and studied long ago as a toy model of an interface in a
random medium \cite{Villain}: one possible interpretation is that $x$ is the coordinate of
a particular point on the interface, which feels a random pinning potential
$E(x)$, while the quadratic
potential is a mean-field description of the elasticity due to the rest of the interface. Another
interpretation (in the context of Bloch walls) is to neglect the deformation of the 
interface, which is only described
by its center of mass coordinate $x$. The quadratic potential is then induced 
by the demagnetizing fields due to the surrounding Bloch walls. 
 
We want to compute  the low temperature properties
of this system in the limit when $\kappa \to 0$. For instance one would like to
know the typical displacement of the ground state, measured through 
$\overline{<x^2>}$,
or the average ground state energy, etc. In
the special case where the energy is Gaussian distributed,
this problem has already been
studied by scaling arguments \cite{Villain}, or with a Gaussian replica variational method \cite{MP2}.
We shall provide hereafter the exact solution, first using a direct extreme value 
statistics approach and then with the replica method.

\subsection {Extreme value approach}

For simplicity, we restrict to the case $d=1$; the extension to higher 
dimensions is however immediate. For temperatures going to zero, we want to find 
the 
minimum of all the energies ${\kappa \over 2 } x^2 +E(x)$ when $x$ scans a 
one dimensional lattice. The joint probability that this minimum is achieved on a
point $x^*$ and takes a value $V(x)={\kappa \over 2 } x^{*2}+E$ is given by:
\be
P(x^*,E)={P(E) } \prod_{x' \ne x^*} 
\((1-P_<\((E+{\kappa \over 2}x^{*2}-{\kappa \over 2} x'^2\)) \))
\ee
For $\kappa \to 0$ we can safely take a continuum limit and we get:
\be
P(x^*,E)={P(E) \over 1-P_<(E)} \exp\((\int dx' 
\log\[[1-P_<\((E+{\kappa \over 2}x^{*2}-{\kappa \over 2} x'^2\]] \))
\))
\label{PxE}
\ee
Integrating over $E$ we get the probability that the minimum
is achieved on point $x^*$. For small $\kappa$, the minimum $E$ is 
expected to be negative and large, and hence only the region where $P_<$
 is small
will be of importance. Rescaling $x^*$ as $x^*= \hat x^*/\sqrt{\kappa}$, 
we obtain
\be
P(\hat x^*) \simeq \int dE  P(E) \exp\((-\int {dz \over \sqrt{\kappa}}
P_<\((E+{{\hat x}^{*2} - z^2 \over 2}\)) \))
\ee
For small $\kappa$ it is thus clear that the relevant energy region is the one around
the value $E_c$ such that $P_<(E_c)= \sqrt{\kappa}$, or:
\be
 E_c = -\(({\log(1/\sqrt{\kappa}) \over B}\))^{1/\delta}
\label{Pinterm1}
\ee
(Notice that the role of the number $M$ of energy levels in the
first section is played here by the length scale $1/\sqrt{\kappa}$, which is 
natural.)
Expanding the energy around $E_c$ as $E=E_c- {\hat x}^{*2}/2+\eps$, we get:
\be
P_<\((E+{{\hat x}^{*2}-z^2 \over 2}\)) \sim \sqrt{\kappa} \exp\((
 \delta B  |E_c|^{\delta -1}(\eps-{z^2 \over 2}) \))
\ee
The integral over $z$ in (\ref{Pinterm1}) is a thus a Gaussian integral. 
We finally get, after a simple integration over $\eps$:
\be
P(\hat x^*) \propto \exp\((- \delta B |E_c|^{\delta-1} {{\hat x}^{*2} \over 2} \))
\label{Phatstar}\ee
Therefore we have shown that the typical distance
to the origin of the point $x^*$ corresponding to a
minimum energy is
\be
\xi= \((\kappa \delta B |E_c|^{\delta-1}\))^{-1/2}= \frac{1}{\sqrt{\kappa \delta}}
\((\log\(({1 \over \sqrt{\kappa}}\))
\))^{1-\delta \over 2 \delta} B^{-{1 \over 2\delta}} 
\ee
and more precisely the distribution of $x^*/\xi$ is a Gaussian of
unit variance \footnote{Notice that for small $\kappa$ we
have $\kappa \xi^2 << |E_c|$ which justifies our expansion
around $E_c$ in the derivation of $P(\hat x^*)$}.

We can also compute the probability distribution of the ground state
energy ${\cal V}^*$ as
\be
P({\cal V}^*)= \int dx \int dE \ \delta\(({\cal V}^*-{\kappa \over 2}x^2-E\)) P(x,E)
\ee
where $P(x,E)$ is given in (\ref{PxE}). The result is the following: introducing the
rescaled energy $u$ as:
\be
{\cal V}^*= E_c + \frac{1}{2 B \delta |E_c|^{\delta-1}} \log \left[\frac{B \delta |E_c|^{\delta-1}}
{2\pi}\right]
+ \frac{u}{B \delta |E_c|^{\delta-1}}\label{V*}
\ee
one finds that $u$ is distributed according to the universal Gumbel distribution, 
Eq. (\ref{gumbel}). In particular, the extremely deep states are exponentially
distributed, as  $\exp [\mu {\cal V}^*/T]$, with $\mu = T B \delta |E_c|^{\delta-1}$.

\subsection{Replica approach}

Interestingly, the replica approach with a one step {\sc rsb} also leads to the 
{\it exact} result. Introducing again the generating function $g(\lambda)$ of
$P(E)$, we have, for large $\lambda$:
\be
g(\lambda)=\frac{\delta -1}B^{-1/\delta-1} \((\frac{\lambda}{\delta}\))^{\delta/\delta-1}
\ee
The average of ${\cal Z}^n$ can thus be expressed as:
\be
{\overline{{\cal Z}^n}} = \sum_{x_1,..x_n}  \exp\left[-\frac{\kappa}{2 T} \sum_{a=1}^n 
x_a^2 + \sum_x g(\frac{1}{T}\sum_{a=1}^n \delta_{x,x_a})\right]
\ee
Let us make the ansatz that at low temperature, the saddle point of this expression 
is 
such that:
\bea
x_1=x_2=...=x_m=y_1\\
x_{m+1}=x_{m+2}=...=x_{2m}=y_2\\
...\\
x_{n-m+1}=...=x_n=y_{n/m}
\eea
and perform the Gaussian integration over the $y_i$. We finally obtain:
\be
{\overline{{\cal Z}^n}} = \exp \frac{n}{m} \left[\frac{1}{2} \log \frac{2\pi T}{\kappa m}
+ g(\frac{m}{T})\right] \equiv \exp -\frac{n}{T} f(\rho)
\ee
with $\rho=m/T$. Looking for the extremum of $f$ as a function of $\rho$ we find, 
in the limit $\kappa \to 0$ (and with $\delta > 1$),
\be
\rho^* = \delta B^{1/\delta} (\log \frac{1}{\sqrt{\kappa}})^{(\delta-1)/\delta} \equiv 
B \delta |E_c|^{\delta-1}\label{rho}
\ee
where $E_c$ is given by Eq. (\ref{Pinterm1}). 

It is easy to check that the free energy $f(\rho^*)$ precisely reproduces the 
above result for the ground state energy obtained directly, Eq. (\ref{V*}).
As explained above, the calculation of the quantities $Y_k$ within the replica 
method indicates that the low energy states are exponentially distributed
with a parameter given by $\rho^*=m/T$. Hence, comparing (\ref{V*}) and
(\ref{rho}), we see that the replica method indeed predicts the correct  statistics
of deep states. The replica method also allows one to calculate 
\be
\overline{P(x)} =\left. \sum_{x_2,..x_n}  \exp\left[-\frac{\kappa}{2 T} \sum_{a=1}^n 
x_a^2 + \sum_x g(\frac{1}{T}\sum_{a=1}^n \delta_{x,x_a})\right] \right|_{x_1=x}
\ee
Within the above one step solution, this immediately leads to the following 
Gaussian result:
\be
\overline{P(x)} = \sqrt{\frac{\kappa \rho^*}{2 \pi}} \exp -\frac{\kappa \rho^* x^2}{2}
\ee
which is identical to Eq. (\ref{Phatstar}).

The replica method also allows one to discuss the non zero temperature 
regime, which is much harder to study directly. As shown above, there is a 
phase transition towards a `delocalised' 
phase where $Y_k \equiv 0$ when $\mu=\rho^* T=1$ \footnote{Note however that there is a true phase transition only in the limit $\kappa \to 0$ or $d \to \infty$, i.e. when the number of degrees of freedom is infinite. Otherwise, the 
transition for $\mu=1$ is really a crossover.}. However, for any non-zero temperature and for $\delta > 1$, the system eventually reaches $\mu=1$ for small enough $\kappa$. This can be interpreted as follows: as $\kappa \to 0$, the number of accessible states diverges. But since the difference between
the ground state and the first excited state decreases as $|E_c|^{1-\delta}$ 
(when $\delta > 1$), it does become smaller than $T$ for a sufficiently small
$\kappa$, beyond which a large number of quasi-degenerate states  
contribute to the partition function, as in the high temperature phase. 
Only for $\delta=1$ is there a true transition temperature, independent 
of $\kappa$ (see \cite{Monthus} for a discussion of this point in a different
context). For $\delta < 1$, one expects a first order phase transition (see 
above).

Finally, let us note that in the case where the confining potential is harmonic
(i.e. equal to $\kappa x^2/2$), the Gaussian {\it variational} replica method
developed in \cite{MP,Engel} also gives the {\it exact} result for $\rho^*$.

\subsection{Physical interpretation of the replica solution}

Using the replica method, one can also compute higher moments of $P(x)$, such as 
$\overline{P(x)P(y)}$, etc. One can then show that the replica solution is identical to
the following probabilistic construction for $P(x)$ for a given sample:
\be
P(x) = \frac{1}{\cal Z} \sum_\alpha w_\alpha \delta_{x,x_\alpha}\label{Pomega}
\ee
where the $w_\alpha$ are random weights, chosen with a probability distribution given by 
Eq. (\ref{Pw}), and the $x_\alpha$ are random variables, independent from the $w$'s, and chosen according to a Gaussian of width $\kappa^{-1}$. 

\section{The random energy model and Burgers' turbulence}

\subsection{The Cole-Hopf transformation}

It is well known that the solution of Burgers equation with a random initial
velocity field can be expressed as a partition sum of the form Eq. (\ref{Z}).
Let us restrict for simplicity to one dimension, although, again, generalisation to 
higher dimensions is possible. The Burgers equation in the absence of forcing reads:
\be
\frac{\partial v}{\partial t}+ v \frac{\partial v}{\partial x} = 
\nu \frac{\partial^2 v}{\partial x^2} 
\ee
where $\nu$ is the viscosity. The initial velocity field $v(x,t=0)$ will be chosen
as $v(x,t=0)=\frac{\partial E(x)}{\partial x}$. Writing $v = -2 \nu \frac{\partial
\log {\cal Z}}{\partial x}$, allows to transform the Burgers equation into the following
linear diffusion equation (Cole-Hopf transform): \be
\frac{\partial {\cal Z}}{\partial t} =  \nu \frac{\partial^2 {\cal Z}}{\partial x^2}
\ee
with initial condition ${\cal Z}(x,t=0)=\exp[-E(x)/2\nu]$. The solution thus reads:
\be
{\cal Z}(x_0,t) = \int_{-\infty}^{+\infty} \frac{dx}{\sqrt{4 \pi \nu t}} \exp\((-\frac{1}{2 \nu} [\frac{(x-x_0)^2}{2 t}
+ {E(x)}] \))\label{Zburgers} 
\ee
which is, up to a multiplicative factor, identical to the `spatial' {\sc rem} 
defined by Eq.(\ref{Z}) with the
following identification:
\be
T \to 2 \nu \qquad \kappa \to \frac{1}{t}
\ee
Physically, the disordered problem associated to this {\sc rem} is that of a point
particle interacting with a (random) pinning potential $E(x)$, attached by a
spring to point $x_0$, which is a simplified model for an 
extended elastic object in a random potential.
 This model was also recently considered in the context of solid friction \cite{Nozieres}. 

Although the two problems, spatial {\sc rem} on one hand and decaying Burgers turbulence
on the other hand, are formally identical, they may differ by the type of questions
one wants to address. For instance in turbulence one is interested in the correlations
of velocities, which involves knowing the variations of the free
energy of the {\sc rem} (\ref{Zburgers}) when $x_0$ varies.
The case where $E(x)$ is random with short range correlations
correspond to  a short range correlated velocity field $v(x,t=0)$ with a `blue'
spectrum (i.e.  $|v(k,t=0)|^2 \propto k^2$, where $k$ is the Fourier variable) and
the small viscosity (large Reynolds) limit corresponds to small temperature in the
associated disordered problem.

\subsection{Cusps and shocks}

\begin{figure}
\centerline{\hbox{\epsfig{figure=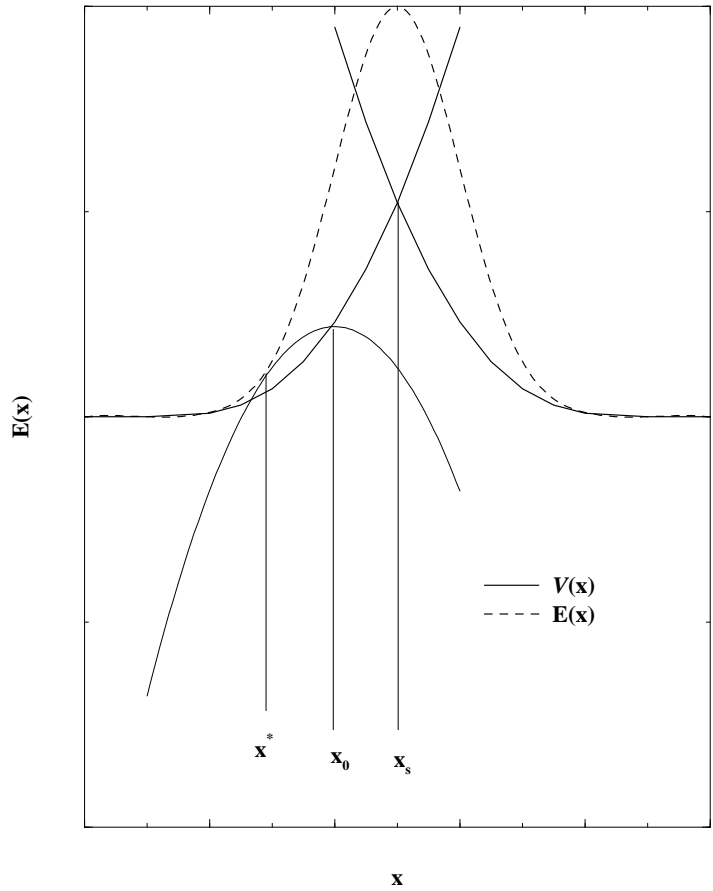,width=8cm}}}
\vskip 0.8cm
\caption{Graphical solution of the Burgers equation in the limit of small 
viscosity, in the neighbourhood of a cusp. 
The dashed line is the original potential $E(x)$, while the full
line corresponds to the effective potential ${\cal V}^*(x)$. The curves actually
continue beyond the cusp of ${\cal V}^*(x)$, where one metastable and one stable
saddle point coexist.}
\label{fig3}
\end{figure}

 In the zero viscosity (or zero temperature) limit, the partition function
(\ref{Zburgers}) can be evaluated by a saddle point method. For a fixed $x_0$, one looks
for the value of $x^*$ such that $\kappa (x_0-x^*)^2/2 + E(x^*)$ is minimum. 
The saddle point
construction \cite{Burgers} is graphically  explained in Fig. (\ref{fig3}), for a simple profile $E(x)$.
 For a given $x_0$, one draws as a function of $x$
 the parabola ${\cal V}- \kappa (x_0-x)^2/2$ and looks for the minimum
value of ${\cal V}$, called ${\cal V}^*(x_0)$, such that this parabola intersects
 the curve $E(x)$; calling $x^*$ the intersection point, the saddle point approximation gives:
 \be
{\cal Z}(x_0,t) \simeq \exp[-\frac{{\cal V}^*(x_0)}{2\nu}] \simeq \exp\((-{1 \over 2 \nu}
[{\kappa \over 2} (x_0-x^*)^2 + E(x^*)] \))
\ee

For large values of $\kappa$, the
parabola is very sharp, and there is only one `optimal' intersection point $x^*$ for
each value of $x_0$; to a first approximation, one thus has ${\cal Z}(x_0,t) \simeq
\exp[-E(x_0)/T]$. 
On the other hand for very small $\kappa$, which corresponds to the large time limit
of the decaying Burgers turbulence, the parabola ${\cal V}-
\kappa (x_0-x)^2/2$ is extremely flat
and the intersection points will be determined by the extreme (negative) values of
the potential $E(x)$. In this limit, the statistics of the effective potential
${\cal V}^*$ -- and thus of the velocity field $v(x,t)$ -- reflects the statistics of
the extreme values of $E(x)$, and is thus, to a large degree, universal. Generically,
the solution $x^*$ depends very weakly on $x_0$ and the effective potential
 ${\cal V}^*(x_0)$ can thus
approximatively be written as:
\be
{\cal V}^*(x_0) \simeq \frac{\kappa}{2} (x_0-x^*)^2 + E(x^*)\label{Vstar}
\ee
with a {\it fixed } $x^*$,
where $E(x^*)$ corresponds to a particularly `deep' minimum $x^*$ of
the potential $E(x)$. This is the generic situation when one varies $x_0$ locally; it corresponds to a velocity field which is locally linear:
\be
v(x_0) = \frac{d{\cal V}^*(x_0)}{dx_0} = \kappa (x_0-x^*)
\ee
(Remember that by definition $\partial {\cal V}^*/\partial x^* =0$). There exist however exceptional values
$x_s$ of $x_0$ such that the first intersection of the parabola and the curve $E(x)$
appears simultaneously at two
points $x_1^*<x_2^*$: when  $x_0$ varies from $x_s-\eps$
to $ x_s+\eps$, the solution $x^*$ jumps from
$x_1^*$ to $x_2^*$. This corresponds to a cusp in the minimum value ${\cal V}^*$ as a
function of $x_0$ (see Fig. (\ref{fig3})). In the language of Burgers' turbulence, this is a
shock since the velocity $v$ (which is the derivative of ${\cal V}^*$) is
discontinuous at $x_0=x_s$.

\subsection{Decay from an uncorrelated $E(x)$ configuration: Kida's analysis}
\label{kida}

Let us now focus onto the case where $E(x)$ is randomly distributed with a short
range correlation, and the time $t$ is large,
corresponding to a very small $\kappa$. This limit was studied in detail by Kida in
the context of Burgers' equation \cite{Kida} (see also \cite{Frisch}). Let
us denote by $x_\alpha$ the various values of the intersection points between 
the parabola and the curve $E(x)$ when one varies $x_0$.
  After a proper coarse graining one can totally forget about the correlations of 
 $E(x)$, and thus  the $\{x_\alpha\}$ are randomly (Poisson) distributed along the 
$x$-axis. If the distribution of $E$ decays as $\exp -B|E|^\delta$, the
extreme value statistics tells us that the distribution of $E_\alpha \equiv E(x_\alpha)$ is of the Gumbel type.
The only delicate point is to understand what is the effective number of
independent variables, $M$, appearing in this distribution.
This number depends on $\kappa$ and is determined
self-consistently as follows: as $x_0$ departs from $x_\alpha$, at some point
(because of the quadratic growing term  $\kappa (x_0-x_\alpha)^2/2$) will a better
saddle point $x_\beta$ be prefered. Since the width of the Gumbel
distribution is given by:
\be
\frac{1}{\delta B^{1/\delta}} (\log M)^{\frac{1-\delta}{\delta}}\label{delta}
\ee
(see Eq. (\ref{ecprime}) above), this sets the order of magnitude of the
difference between $E_\alpha$ and $E_\beta$, which must also be, by definition, of
the order of $\kappa(x_\alpha-x_\beta)^2$. Furthermore, taking the correlation length of the potential $E(x)$ to be $1$, the effective number
of independent variables is given by:
\be
M = |x_\beta-x_\alpha| \simeq \frac{1}{\sqrt{\kappa \delta B^{1/\delta}}}
(\log M)^{\frac{1-\delta}{2\delta}}\ee
or, to logarithmic accuracy, and using the correspondance $\kappa \to 1/t$,
\be
M \propto \sqrt{t} \left(\log t \right)^{\frac{1-\delta}{2\delta}}
\label{Mscale}
\ee
Note that by definition, $M$ is also the typical distance beween two shocks $\ell(t)$, which is
thus seen to grow as $t^{1/2}$ with logarithmic corrections (these corrections
disappear for $\delta=1$, where the initial potential already possesses the
universal exponential tail).
 This is one of the important
results of the original analysis of Kida. Furthermore, since the local 
slope of the velocity is $\kappa=1/t$, the maximum velocity is of order:
\be
v_{\max} = \frac{\ell(t)}{t} \simeq \frac{\left(\log t \right)^{\frac{1-\delta}{2\delta}}}{\sqrt{t}}
\ee
which corresponds to a time dependent Reynolds number :
\be
Re = \frac{v_{\rm max} \ell}{\nu} \propto \frac{\left(\log t \right)^{\frac{1-\delta}{\delta}}}{\nu}
\ee
which goes to zero (albeit very slowly) when $t \to \infty$ for $\delta > 1$.
This is similar to the above remark that for any small temperature, the system
goes back into its high temperature phase when $\kappa \to 0$.

Using this construction, and the full distribution of the $E_\alpha$, Kida was able to
obtain directly the large time behaviour of the two point velocity correlation,
 $\overline{v(x) v(x+r)}$,
which is a universal function once the lengths are expressed in terms of the mean
distance between two shocks $\ell(t)$, and the velocities in units of $v_{\max}$ \cite{Kida}. His result is recalled in
the appendix. Let us  show  how one can obtain precisely the same results using the
replica method,
which in fact provides the ful probability distribution function of $v(x+r)-v(x)$. 

\subsection{The replica analysis}

Let us first note that Eq. (\ref{Vstar}) can alternatively (in the limit $\kappa, T \to 0$) be written as an infinite sum:
\be
{\cal Z}(x_0) = \exp[-{\cal V}^*(x_0)/T] = \sum_\alpha w_\alpha \exp[-{\kappa(x_0-x_\alpha)^2
\over 2 T}],
\label{Zomega}
\ee 
where $x_\alpha$ are Poisson distributed with an arbitrary (see below) linear density $\sigma$. The $w_\alpha$ are independent random variables again chosen according to the distribution (\ref{Pw}), with 
$\mu$ given by (see Eq. (\ref{mumu})):
\be
\mu = T {\delta B^{1/\delta}} (\log M)^{-\frac{1-\delta}{\delta}}
\label{murep}
\ee
That Eq. (\ref{Zomega}) precisely reproduces Kida's construction comes from the fact that, as $T \to 0$, the distribution of weights becomes so broad that the sum determining
${\cal Z}(x_0)$ becomes entirely dominated by a single term, which is the one which maximizes
$w_\alpha \exp[-\kappa(x_0-x_\alpha)^2/2]$. Again, the corresponding $x_\alpha$ switches
discontinuously as a function of $x_0$, when another value $x_\beta$ suddenly takes over. 
This construction is independent of the density $\sigma$, provided that $\sigma M \gg 1$ (i.e., in the long time limit). 

The crucial point now is that the explicit construction (\ref{Zomega}) actually 
gives results which are {\it identical} to those obtained using a replica representation:
\be
\overline{{\cal Z}(x_1){\cal Z}(x_2)...{\cal Z}(x_n)} = \sum_{\pi} \exp[\frac{1}{2} \sum_{a,b=1}^n R_{\pi(a),\pi(b)} x_a x_b]\label{Zreplicas}
\ee
in the limit $n \to 0$. In the above expression, $\pi$ denotes a permutation of the $n$
replica indices, and 
 the $R_{ab}$ matrix is a one step {\sc rsb} matrix \cite{MPV} with elements
 $R_{ab}=r$ when $a$ and $b$ are in the same diagonal
block of size $m$, 
$R_{ab}=0$ when $a$ and $b$ are in different blocks, and $R_{aa}={(1-m) r}$,
enforcing the sum rule $\sum_{b=1}^n R_{ab} =0$. By equivalent results we mean
that one can compute quantities like the average probability of
being in $x$,
\be
\overline{P(x)} = \overline{\frac{{\cal Z}(x)}{\int dx' {\cal Z}(x')}} \equiv \left.\int dx_2...dx_n \overline{{\cal Z}(x){\cal Z}(x_2)...{\cal Z}(x_n)}\right|_{n=0}
\ee
or the average of the product of the two probabilities to be in $x$ and $y$,
\be
\overline{P(x)P(y)} = \overline{\frac{{\cal Z}(x){\cal Z}(y)}{(\int dx' {\cal Z}(x'))^2}} \equiv \left.\int dx_3...dx_n \overline{{\cal Z}(x){\cal Z}(y){\cal 
Z}(x_3)...{\cal Z}(x_n)}\right|_{n=0}
\ee
or generalizations thereof, by both methods (here the average is taken over the realization of
the initial velocity profile). It has been shown in \cite{BMP} (and we recall the main steps of
the derivation in the appendix) that the velocity correlation function 
$\overline{v(x,t)v(y,t)}$ can be computed either directly from Eq. (\ref{Zomega}), or using the representation (\ref{Zreplicas}). The important points are the following:

-- After a proper choice of length and velocity scales, the 
$\overline{v(x,t)v(y,t)}$ correlation function is indeed {\it identical} to the 
result obtained by Kida (see appendix), which is 
the consequence of the fact that both approaches actually rely on the universal 
structure of the extreme events which control the velocity field for large times.

-- The present formalism allows us to extend Kida's results in several directions.
 For example,  the full probability distribution function of $v(x)-v(y)$ has
been computed in \cite{BMP}.
 The problem of decaying Burgers turbulence in higher dimensions can also 
be addressed.

-- The presence of shocks, which manifests itself as a $|x-y|$ singularity in $\overline{v(x,t)v(y,t)}$ at short distances, is intimately connected with the
breaking of replica symmetry \cite{BMP,BBM}: for a replica symmetric matrix $R_{aa}=\tilde R$, $R_{a \neq b} = R_1$, $\overline{v(x,t)v(y,t)}$ is {\it regular} 
for $x \to y$.  As discussed above, these shocks reflect, in
the associated disordered problem, the existence of some {\it metastability} (see Fig. (\ref{fig3}). From a technical point of view, it is interesting to see on this example how metastability is
associated to {\sc rsb} and, as emphasized in \cite{BBM}, to the existence of a short-distance singularity in the effective free-energy ${\cal V}^*$. Precisely the same behaviour is obtained via the Functional Renormalisation Group ({\sc frg}) \cite{FRG}: a singularity appears in the renormalized correlation function of the effective
free energy at scales larger than the `Larkin length', which is the scale
beyond which metastability effects become important. (However, the way to 
handle the shocks correctly within the {\sc frg} is still an open problem
\cite{BBM}).

\section{Perspectives and other universality classes}

As is the case for the central limit theorem, there are other universality classes, distinct from the Gaussian, when one relaxes the hypothesis of a finite variance or of independent 
variables (or both) \cite{BG}. This is also true for the statistics of extremes, and it is interesting to discuss how this might translate into a replica language. Two main 
directions can be thought of: independent variables with other
types of distributions, or correlated variables.

Let us first 
consider the case where the energy levels $E_i$ are still independent, but with a 
tail for large negative $E$ decaying as a power-law, $|E|^{-1-\delta}$. In this case, the 
extreme values are distributed according to the so-called Fr\'echet distribution, which is different from the Gumbel distribution (for example, it decays asymptotically as a power-law with the same exponent $\delta$). Rescaling $E$ by $M^{1/\delta}$ to keep the gap between the ground state and first
excited state finite as $M \to \infty$, one can calculate the quantities $Y_k$ defined in (\ref{Ykdirect}). One finds, for $M$ large but finite:
\be
Y_k = 1 - \exp -\left(\frac{1}{T \log M}\right)^\delta  \qquad (\delta < 1)
\ee
{\it independently} of $k$. (Similar results are obtained for $\delta > 1$).
This is clearly different form Eq. (\ref{Ykdirect}). Notice that this case cannot be addressed within the 
replica method without
some modifications since all the positive moments $\overline{{\cal Z}^n}, \ n>0$ diverge !

Another universality class corresponds to $E_i$ which are strictly bounded, i.e.
$E_i=E_0+\epsilon$, with $\epsilon \geq 0$. More precisely, the distribution of $\epsilon$
for $\epsilon \to 0$ is of the form $P(\epsilon) = \epsilon^{\delta}$ for $\epsilon$ small.
 The resulting distribution of extremes is then called the Weibull distribution.
Rescaling the energies by a factor $M^{\delta+1}$, we find through a direct computation that
$Y_k$ is non trivial for all temperatures, i.e., the model is always in a low temperature phase. For instance in the case $\delta=0$ one gets
 $Y_k(T \to \infty) \sim k T^{1-k}$, and 
$Y_k \sim 1-{\boldmath C}T$ for $T \to 0$ ($\boldmath C$ is Euler's constant).
This is again clearly different from the 1 step {\sc rsb} result (\ref{Ykdirect}). One might hope
that such a situation will lead to a new type of {\sc rsb}, but the situation seems
more complicated.  In the particular case $\delta=0$, one finds:
\be
\overline{{\cal Z}^n} = \sum_{i_1,...i_n} \prod'_i \frac{T}{\sum_{a=1}^n \delta_{i,i_a}}
\ee
where the product is only over the sites $i$ such that $\sum_{a=1}^n \delta_{i,i_a} >0$. The entropy of the replica symmetric solution becomes 
negative below the temperature $T_c=1/e$. Assuming a one step {\sc rsb} saddle point for $T < T_c$ leads to a constant average energy, equal to $1/e$; however, the true ground state energy can be calculated directly and is equal to
$1$. Again, this result is not of the replica type. In this case however, a sensible replica calculation can be 
undertaken since all the $\overline{{\cal Z}^n}, \ n>0$ are convergent. The problem now is that the free-energy cannot 
be calculated by a {\it saddle point} method: replica fluctuations are always important.

In any case, from the point of view of Burgers' turbulence, it is interesting to notice
 that initial 
conditions for the velocity field which do not belong to the exponential universality class
considered by Kida will lead to rather different flow structures at long times, even
within the class  of $E(x)$ functions with local correlations.

Let us now turn to the case where the $E_i$ are Gaussian but long-range correlated; for 
example the case where $E$ depends on a $d$-dimensional space variable $\vec x$, and such that:
\be
\overline{\tilde E(\vec q) \tilde E(\vec q')} = {\delta(\vec q+\vec q') \over q^{2-\eta}}
\ee
leading to
\be
\overline{(E(\vec x)-E(\vec y))^2} \propto |x-y|^{\max(0,2-d-\eta)}
\label{ecorfou}
\ee
The case $\eta=0$, $d=1$ corresponds to a random walk for $E(x)$, which has 
been studied in detail both in the context of Burgers' turbulence \cite{Burgers}, and also as a partly solvable spatial {\sc rem} \cite{Villain,Comtet,Broderix}. The general $\eta,\ d$ case 
has not been solved yet. It has been studied by the
 Gaussian variational replica formalism of \cite{MP}, which shows \cite{MP2} that the case $\eta < 2-d$ (corresponding to a growing correlation function (\ref{ecorfou})) requires `continuous' {\sc rsb}, while the case $\eta > 2-d$
only requires a `one step' breaking. Independent variables correspond to $\eta=2$, i.e. a white spectrum for $E(\vec q)$. We conjecture 
here that the case $\eta > 2-d$ belongs to the same (one step {\sc rsb}) universality class as the {\sc rem} ($\eta=2$). It is actually not difficult to  show directly that the quantities:
\be
c_n=\frac{\overline{{\cal Z}^n}-{\overline{\cal Z}}^n}{{\overline{\cal Z}}^n}
\qquad n=2,3,..
\ee
diverge with the system size below a certain $n$ dependent critical temperature
which is independent of $\eta$ for $\eta > 2-d$, and identical to those found
in the {\sc rem}. This suggests that the one step solution indeed remains
{\it exact} for all $\eta > 2-D$. Preliminary numerical simulations \cite{Andrea} seem to confirm this. This points towards a rather natural result, namely the fact that {\it  weak enough correlations} (measured here by $2-\eta$) between the random variables {\it does not change the universality class for the extreme value
statistics}. This is actually a Theorem for the case $d=1$: for all $\eta > 1$, the extreme value statistics is indeed of the Gumbel type \cite{Galambos2}, while some corrections appear in the marginal case $\eta=1$. Interestingly, a related conjecture was proposed recently for the $d=2$ problem with $\eta=0$, which corresponds to the localization of
electrons in a random magnetic field \cite{Chamon}. In this two dimensional case, the choice $\eta=2-d=0$ corresponds to a marginal logarithmic growth of 
the correlations. 

Returning to the one dimensional case, the situation changes drastically
 when $\eta < 1$, which corresponds to  a typically `growing' profile $E(x)$. The ratios $c_n$ diverge with the system
size for all temperatures, suggesting indeed a change of universality class.
The only known possibility at present is then to describe the system within a  
`continuous' {\sc rsb}, which can be interpreted as a recursive tree-like construction of the low-lying energy state.  In particular, the correlation of the low-lying states have a well
known ultrametric structure. How well this ultrametric structure (known to be exact for
the case where the dimension of $\vec x$ is infinite) reproduces the  distribution and correlations  of the low lying states in finite dimension 
is an open problem \footnote{Note that the average ground state energy predicted by the Gaussian {\it variationnal} replica theory does not lead back to the exact result \cite{Broderix} in the soluble random walk case $\eta=0$.}. We leave this problem for further studies \cite{Andrea}.

In summary, we have argued for the one dimensional problem 
 that the one step replica symmetry breaking scheme
encodes exactly the results on the statistics of extremes for variables
which are: 

i) not too correlated (i.e. when the above exponent $\eta$ is larger than $2-d$)

ii) distributed asymptotically as generalized exponentials (i.e. as $\exp -|E|^\delta$).

The case of long range correlations may correspond, in some limit, to the
`continuous' {\sc rsb} scheme. However, the precise link between the two
is not clear to us at present, and we think that it is an important path
to explore further. It would also be interesting to think of the spin-glass problem from the point of view of the classification of very low energy states
and excitations, which could perhaps provide a natural link between replicas and
`droplet-like' descriptions \cite{FH}.   

\vskip 1cm
{\sc acknowledgments} We thank A. Baldassarri, R. Cont and V. Dotsenko
for many interesting discussions. MM thanks the SPhT in the CEA Saclay for
its hospitality.

\def\cM{{\cal M}}

\vskip 1cm
{\bf Appendix}
\vskip 0.5cm
In this appendix we explain briefly how the replica method and the
direct probabilistic analysis lead to the same result
for the two point correlation
function in decaying Burgers turbulence at large
times. We are interested in the case where
$E(x)$  has local correlations (see  section \ref{kida}), in
which case the result of Kida reads:

\be 
\overline{v(x) v(x+r)} = v_{\max}^2 H\(({r \over \ell(t)}\))
\ee
where  $\ell(t)$ is the mean distance between two shocks, $v_{\max}=\ell(t)/t$, and 
\be
H(x) \equiv {1 \over \sqrt{2 \pi}} {d \over dx} x \int_0^\infty { dy \over \phi(x+y)+\phi(x-y)}
\label{vvKida}
\ee
and $\phi$ is an error function: $\phi(x)= \int_0^\infty dz \exp(-z^2+\sqrt{\pi/2} \ x z)$. 

We shall sketch how
these results can be obtained from the replica representation (\ref{Zreplicas}).
The computations are lengthy and already contained in some
previous papers. Here we just want to help the interested reader
to find his way in the litterature in order to obtain the result.
One starts 
from the replicated partition function (\ref{Zreplicas}):
\be
\overline{{\cal Z}(x_1){\cal Z}(x_2)...{\cal Z}(x_n)} = \sum_{\pi} \exp[\frac{1}{2} \sum_{a,b=1}^n R_{\pi(a),\pi(b)} x_a x_b] \ , \label{apZn}
\ee
where the $R_{ab}$ matrix is a one step {\sc rsb} matrix \cite{MPV} with elements
 $R_{ab}=r$ when $a$ and $b$ are in the same diagonal
block of size $m$, 
$R_{ab}=0$ when $a$ and $b$ are in different blocks, and $R_{aa}={(1-m) r}$,
enforcing the sum rule $\sum_{b=1}^n R_{ab} =0$.
The first step, derived in the appendix D of \cite{BMP},  deduces from (\ref{apZn})
the correlation between the powers $n/2$ of the partition function:
\be
\overline{Z(x,t)^{n/2} Z(y,t)^{n/2}}=
 {2^{n} \over B(-n/2,-n/2)}
\int_0^\infty {d\mu \over \mu} \mu^{-n/2} 
\[[{m\sqrt{r_1} \over 2 \pi} \int dz \((e_x + \mu e_y \))^m
\]]^{n/m} \ .
\ee
where we have defined 
\be
e_x \equiv e^{-m r (z-x)^2/2} \ \ , \ \ e_y=e^{-m r (z-y)^2/2}
\ee
Using the general link $v=-2 \nu {\partial \log {\cal Z} \over \partial x}$,
one gets:
\be
\overline{v(x)v(y)}=\lim_{n \to 0} {16 \nu^2 \over n^2} 
{\partial^2 \over \partial x \partial y}
\overline{Z(x,t)^{n/2} Z(y,t)^{n/2}}=\(({2 \nu m r}\))^2  (g_{11}+g_{12})
\label{vv1}
\ee
where we have defined (the notations are those 
of  \cite{BMP} -Appendix B):
\be
g_{11}=(1-m)
 \int_0^\infty d\mu \ { \int dz \((e_x+\mu e_y\))^{m-2} (x-z)e_x (y-z) e_y
\over \int dz \((e_x+\mu e_y\))^m }
\ee
and:
\be
g_{12}=m
 \int_0^\infty d\mu \ { \((\int dz \((e_x+\mu e_y\))^{m-1} (x-z)e_x \))
 \ \(( \int dz \((e_x+\mu e_y\))^{m-1} (y-z) e_y \))
\over \int dz \((e_x+\mu e_y\))^m } \ .
\ee
This expression could also be derived directly without replicas from
the infinite sum (\ref{Zomega}),
with the identification $m=\mu$, $ r = \kappa/(T \mu)$ as can be
seen from formulas (B10,B11) of \cite{BMP} (where the number $1/(m r)$
was called $\delta$). 

The whole problem is now to 
evaluate this expression in the limit of large Reynolds, which means small $\mu$
or low $T$. In this regime, using the fact that $m$ scales linearly with $T$
and $r$ scales as $1/T^2$, it has been shown in the appendix
B of \cite{BMP} that expression (\ref{vv1}) reduces to:
\bea \nonumber
\overline{v(x)v(y)}& =&
{\kappa T \over \mu} 
\(( 
\sqrt{{2\over \pi}} \int_0^\infty dh
{
   e^{-h^2/2} \ [ h^2-d^2/4 ] 
\over 
   e^{d^2/8} \[[ e^{-h d/2} \cM_0(h-{d \over 2}) + e^{h d/2} \cM_0(-h-{d \over 2}) \]] 
} 
\right.
\\
&-&
\left.
{d\over \pi} \int_0^\infty dh
{
    e^{-h^2}
\over 
    e^{d^2/4} \[[
    e^{-h d/2} \cM_0(h-{d \over 2}) + e^{h d/2} \cM_0(-h-{d \over 2})\]]^2
} 
\)) 
\label{vv_rep}
\eea
where 
\be
\cM_0(x)=  \int_x^\infty {dz \over \sqrt{2\pi}}  e^{-z^2/2} \ \ ; \ \ 
d= {|x-y| \over \sqrt{T/(\mu \kappa)} }
\ee
So the natural length scale appearing
in this solution is $\ell= \sqrt{T/(\mu \kappa)}$. Using (\ref{murep},\ref{Mscale}),
one sees easily that it precisely scales 
at large times as the average distance between
shocks of Kida's analysis.
In terms of reduced lengths, one can check that the two distributions
(\ref{vv_rep}) and (\ref{vvKida}) are actually identical.

\end{document}